# BAYESIAN NETWORK BASED XP PROCESS MODELLING

Mohamed Abouelela, Luigi Benedicenti

Software System Engineering, University of Regina, Regina, Canada

## ABSTRACT

*A Bayesian Network based mathematical model has been used for modelling Extreme Programming software development process. The model is capable of predicting the expected finish time and the expected defect rate for each XP release. Therefore, it can be used to determine the success/failure of any XP Project. The model takes into account the effect of three XP practices, namely: Pair Programming, Test Driven Development and Onsite Customer practices. The model's predictions were validated against two case studies. Results show the precision of our model especially in predicting the project finish time.*

## 1. INTRODUCTION

Extreme Programming (XP) is a lightweight software development methodology. XP is one of the iterative informal development methodologies known as Agile methods. XP comprises a number of values, practices and principles. There is no large requirements and design documents. XP uses what is called User Stories instead of requirements. The XP project comprises of a number of User Stories. Each user stories contains a number of Story Points. The development process constructed from iterative small releases. In each release, User Stories are selected to be developed in this release according to their importance.

Managers of XP projects suffer from lack of prediction systems capable of estimating the expected effort and quality of the software development process. Managers need to know the probability of success or failure of XP project. Models capable of predicting the project finish time are very helpful to the project managers. Those models should also be capable of predicting the product quality in terms of the expected number of defects. These requirements should be covered in strong mathematical model.

In this paper, a Bayesian Network based mathematical model for XP process is presented. The proposed model satisfies the following features:

-      It considers the iterative nature of XP by modelling the project as a number of sequential releases.

-      The model able to predict the expected finish time, and therefore it could determine the success/failure of the project.

-      The prediction can be done in the project planning phase before starting the actual development using very simple input data.

-      The model tracks the developer velocity (measured in number of Story Points per day) as function of the developer experience. It also models the increase in the developer velocity as the project goes on.

-      The model considers the effect of the Pair Programming and Test Driven Development practices on the Team velocity.

-      The model predicts the process quality by measuring the defect rate in each release.

-      It considers the effect of the Onsite Customer and Test Driven Development practices on the defect rate.

The proposed model was implemented using AgenaRisk toolset [1]; a toolset for modelling risk and making predictions based on Bayesian Network. Two case studies were used for the validation of our model. Results show the precision of our model especially in predicting the project finish time.

This paper is organized as follows: in the next section, a survey of the related work and an overview of the Bayesian Network will be provided. Model Design is illustrated in section 3, while the validation is provided in section 4. Finally, conclusions are offered in the last section.

## 2. BACKGROUND

In the next section, a literature survey of related work will be provided. The survey covers a number of the most important XP process models existing in the literature. Then, the Impact of XP Practices on Software Productivity and Quality is illustrated in the following section. Finally, an overview of the Bayesian Network will be provided in the last section.

### 2.1.  Related Work

Although the widespread usage of XP in both academic and industry, only few attempts for modelling XP exists. In this section, we will provide a survey of the most important XP Process models.

In [2], a simulation model was developed to analyse the effects of almost all XP practices on the software development effort. The developed model was applied for a typical XP project and the effects of all the individual practices were calculated. The results showed a reduction in software development cost as increasing the usage levels of all XP practices.

In [3], the authors built a software effort prediction model for XP based on Bayesian networks. The proposed model can learn from project data in order to make quantitative effort predictions and risk assessments. The model has been validated by applying to a real industrial project. Collecting data from the early part of the project enabled the model to update its parameters and improve its predictions. The model could successfully achieve extremely accurate predictions about the level of functionality delivered over time.

In [4], the authors introduced an XP process model to evaluate the effectiveness of XP key practices (Pair Programming, Test-First Programming), and to investigate how the practices influence the evolution of a certain project. To achieve this, software process simulation has been chosen. A process model has been developed and a simulation executive has been implemented to enable simulation of XP software development activities to simulate how the modeled project entities evolve as a result.

Williams and Erdogmus [5] developed a Net Present Value (NPV) model of Pair Programming (PP). The model combines the productivity rates, code production rates, defect insertion rates, and defect removal rates. The authors conclude that PP is a "viable alternative to individual programming."

In [6], the authors model, simulate and analyze the pair programming and pair switching practices. The model explores many variables affecting pair programming efficiency. The results showed that XP efficiency increases with both psychological compatibility and pair adaptation speed between the members of the pair. In addition, the XP process appears to have an advantage over the traditional approach when pair switches are not too frequent.

## 2.2. The Impact of XP Practices on Software Productivity and Quality

A number of Studies were conducted to assess the impact of XP practices such as Pair Programming, Test Driven Development and Onsite Customer practices in the software Productivity and quality. Those contributions study the impact of such practices on the Project Velocity and the product defect rates.

A number of quantitative studies conducted to assess the validity and efficiency of the Pair Programming practice. Some of these contributions study the impact of the Pair Programming practice on the Project velocity. Generally speaking, According to those studies, Pair Programming increases the Project Velocity by a factor starting from 0 (no change) to 45%.

In [7], the authors conducted found that pair programming is 40-50% faster than solo programming. In addition, the pairs implement the same functionality with 20% fewer line of codes than the solo students. In [8], The results showed that there is no significant difference in the average development time between XP with pair programming and XP without Pair Programming. In [9], the authors found that the PP group spends 19% less time than individuals to complete the same project. In [10], the authors found that Pair Programming does not improve neither the quality nor the productivity.

A number of quantitative studies conducted to assess efficiency of the Test Driven Development practice. Some of these contributions study the impact of this practice on the Project velocity and the product defect rate. Generally speaking, According to those studies, Test Driven Development results in increasing the project time by a factor starting from 0 (no increase) in some studies to 80% in others. On the other hand, Test Driven Development results in reducing the defect rate by a factor of 40% in some contributions.

In [11], the authors reported an improvement in quality achieved by a team following test-during-coding process ranged from 38% to 267% fewer defects. Unfortunately, test-during-coding process increase the development time by a factor from 60% to 100%. In [12], the authors reported that the TDD developers took more time (16%) than those not using this practice, but produced higher code quality by a factor of 18%.

A number of studies consider the impact of the Onsite Customer Practice. In [13], the authors reported a reduction in the development effort by a percentage of 5.48%. In [14] four case studies with different degrees of customer interaction were considered. The authors measured the effort spent to fixing defects in the four cases. In the very high customer involvement case, only 6 percent of the effort was spent to fixing defects. However, in the low level of customer involvement, the time spent fixing defects reached about 40 percent.

## 2.3.    Bayesian Network Modelling

Bayesian Network (BN) [15] is a probabilistic graphical model that represents a set of random variables and their conditional probabilities via a directed acyclic graph (DAG). In these graphical structures, each node in the graph represents a random variable, while the edges between the nodes represent probabilistic dependencies among the corresponding random variables. In addition, in a BN each node has an associated probability table, called the Node Probability Table (NPT). These conditional dependencies in the graph are often estimated by using known statistical and computational methods.

Bayesian Networks are used for reasoning and decision-making in problems that involve uncertainty and probabilistic reasoning. The first working applications of BNs (during the period 1988-1995) tended to focus on classical diagnostic problems, primarily in medicine and fault diagnosis [16]. Nevertheless, the medical and biological/DNA decision-support domain has continued to be the most preferred area for published BN applications. On the other hand, companies such as Microsoft and Hewlett-Packard have used BNs for fault diagnosis [16]. The ever-increasing need for improved decision support in critical systems has resulted in a range of BN-based systems. These include BN models for air traffic management, railway safety assessment, and terrorist threat assessment [16]. There have also been numerous uses of BNs in military applications.

The structure of a DAG is defined by two sets: the set of nodes (vertices) and the set of directed edges. The nodes represent random variables, while the edges represent direct dependence among the variables and are drawn by arrows between nodes. In particular, an edge from node $X_i$ to node $X_j$ represents a statistical dependence between the corresponding variables. Thus, the arrow indicates that a value taken by variable $X_j$ depends on the value taken by variable $X_i$, or roughly speaking that variable $X_i$ "influences" $X_j$. Node $X_i$ is then referred to as a parent of $X_j$ and, similarly, $X_j$ is referred to as the child of $X_i$. If $X_i$ has no parents, its local probability distribution is said to be *unconditional*, otherwise it is *conditional*.

Figure 1 shows an example of BN. This graph is part of the proposed XP process model. The number of defected story points depends on both the test driven development practice usage and the defect rate random variables, while the team velocity is affected by only the test driven development practice usage.

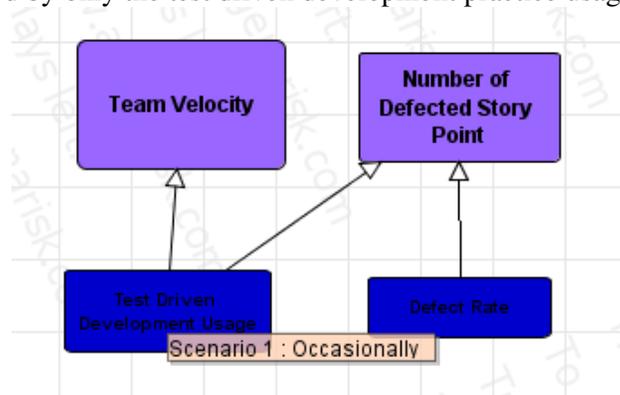

Figure 1 Example for Bayesian Network

## 3. MODEL DESIGN

The proposed model complies XP process iterative structure. The model considers the iterative nature of XP by modelling the process as a number of sequential releases. An example of releases connectivity is shown in Figure 2. Each release has eight inputs and two outputs. The output of one release is considered as an input for the following release.

In the next section, the model details explaining different inputs and outputs are provided. Then, the details of the Team Velocity Model and Project Defects Model will be provides in the following sections.

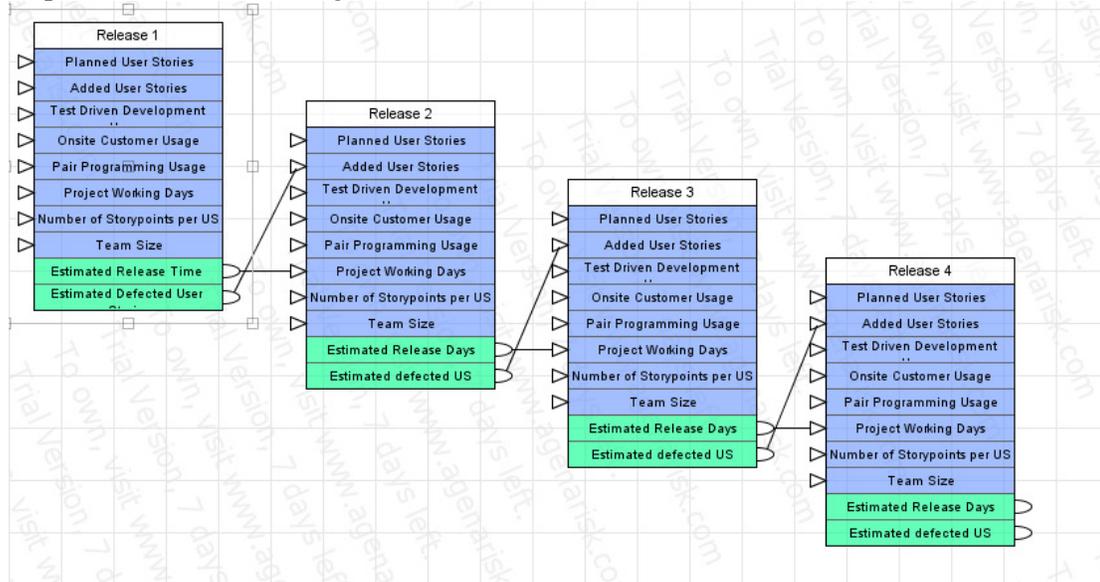

Figure 2 Four releases XP project Model

### 3.1. Model Overview

Figure 3 shows the based components of the proposed model. The model comprises eight input parameters, two estimated output parameters, and two internal models. According to the input parameters values, with the aid of the two internal models, the values of the output parameters (Estimated Release Time and Estimated defected User Stories) can be predicted. In addition, the model considers three XP basic activities: Release Planning, Development Session and Acceptance Test Activities (shown in the elliptical shape). For simplicity, the model assumes the following assumptions:

- Defects are only found at the acceptance test phase.

- The defects are modelled as story points to be treated in the next release.

- The number of defects is affected by *Test Driven Development* and *On site customer* Practices.

- The developer velocity increases as the project goes on.

- The estimated release time is calculated as the time of the development session activity only ignoring the release plan and acceptance test times.

- The team velocity is affected by the team size, team experience, *Pair Programming* Practice, and *Test Driven Development* Practice.

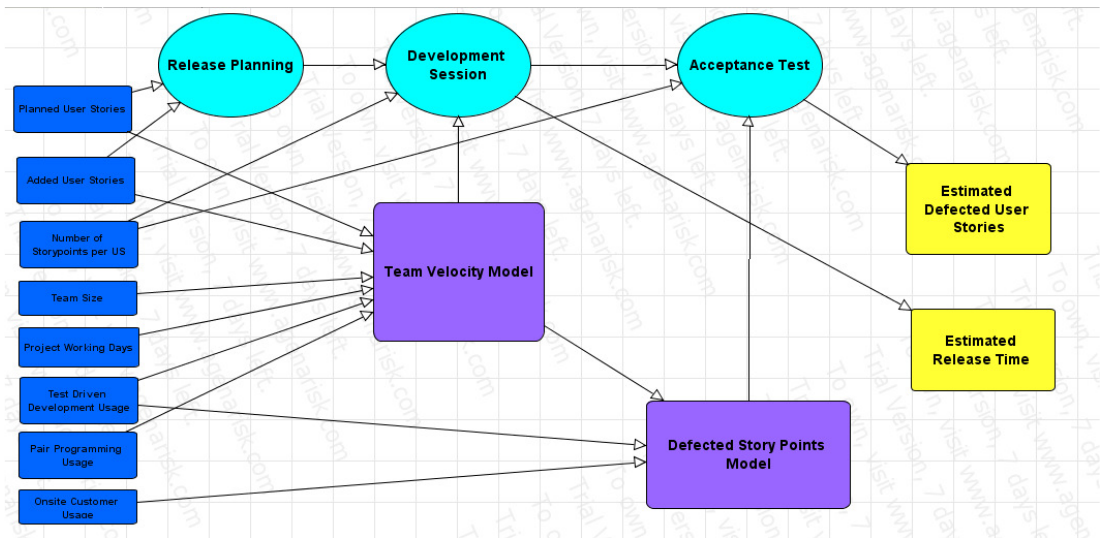

Figure 3 One Release Model basic components

## *Model Input Parameters*

Eight parameters were considered as input parameters to the release model:

- *Planned User Stories*: The number of user stories to be developed in this release. This number should be available before the beginning of the release.

- *Added User Stories*: The number of user stories to be added from the previous release due to the defects in the previous release.

- *Average number of Story Points per User Story*: This number is an estimate number. It can be calculated as an average of the previous releases and similar projects.

- *Team Size*: The number of developers in the development team.

- *Project Working Days*: The summation of the estimated release days over all previous releases.

- *Pair Programming usage*: a scale of 5 levels describing at what extend the team adopts the Pair programming practice is used for measuring. A mapping for the scale to a percentage is done according to table 1.

- *Test Driven Development usage*: The same five levels scale is used to describe at what extend the team adopt the Test Driven Development practice.

- *On Site Customer usage*: The same five levels scale is used to describe at what extend the team adopt the Onsite Customer practice.

Table 1  Scale to Percentage mapping for XP practices usage

| Level of Usage | Equivalent percentage |
|---|---|
| *Never* | 0 |
| *Occasionally* | 0.25 |
| *About half* | 0.5 |
| *Frequently* | 0.75 |
| *Almost used* | 1 |

According to the input parameters values, with the aid of *team Velocity* and *defected story points* Models, the values of the output parameters (Estimated Release Time and Estimated defected User Stories) can be predicted. Those values are feed as an input to the next release. The details of the *team Velocity* and *defected story points* Models will be provided in the next section.

The three basic activities of XP release: Release Planning, Development Session and Acceptance Test are shown in the elliptical shape. In the Release Planning phase, the user stories are sorted according to the importance and the release user stories are selected among them. The development session includes the basic development activities: simple design, coding and unit testing activities. Finally, the release testing activity is done in the Acceptance Test phase.

## 3.2.   Team Velocity Model

Team velocity, measured in number of *user stories* per day, is a good representation for the productivity of the team. The model is shown in Figure 4. Three main factors affect the team velocity in XP, namely: Team XP skills, Pair Programming usage and Test Driven Development usage.

The main components of the model are:

- Developer initial skills: random number uniformly distributed over the range [1,10] [4]. This number represents the developer ability to develop software.

- Developer skills: Dev. Skills depend on two main factor, namely: *dev_initial_Skills* and *project_working_days*. This value represents increasing developer skills with time. It can be calculated as described in Equation (1)[4], where *LC* is the learning coefficient, typically 0.009 [4].

   *Dev_skills = log(dev_initial_Skills + project_working_days*LC)*          Equation (1)

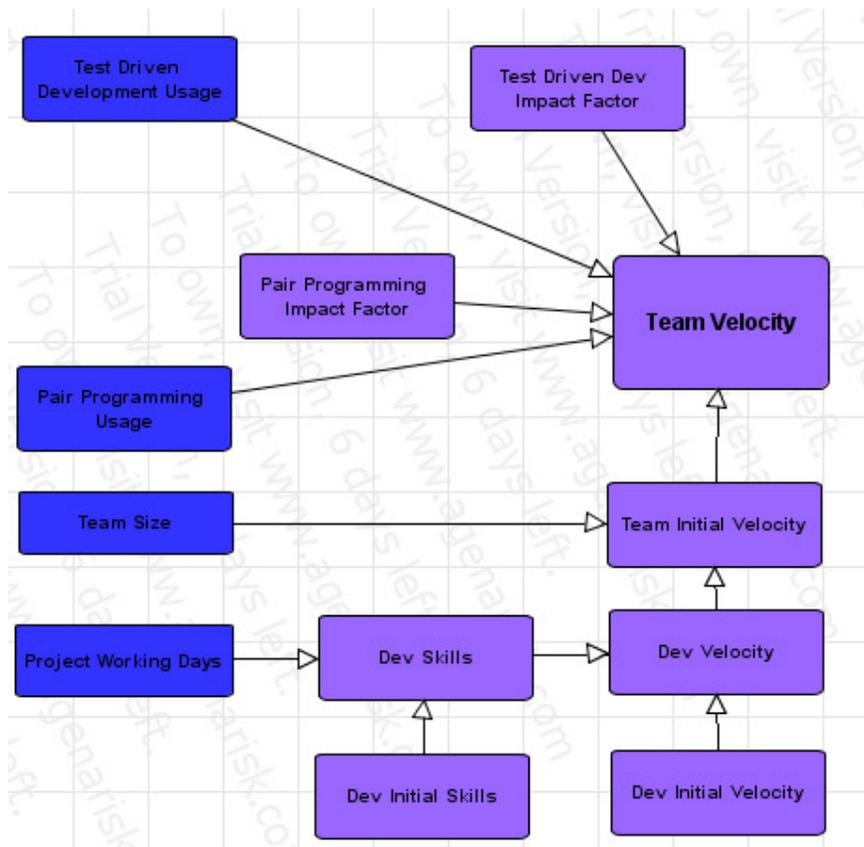

Figure 4 Team Velocity Model

- Developer Initial Velocity: represents the number of story points the user can develop per day. According to empirical data [4], a random variable following the normal distribution with mean 4 and standard deviation 1 was considered to represent this value.

- Dev. Velocity: this value represents the impact of the developer skills on his velocity. It is calculated as the summation of the *Dev. Initial Velocity* and the *dev. Skills* factor.

- Team Initial Velocity: represents the number of story points the team can develop per day. It can be calculated as the product of *Team Size* times *Dev. Velocity*.

- Pair Programming Impact Factor (*pp_Impactfactor*): This factor represents the impact of adopting Pair Programming practices in the team velocity. This value was set to a normal distribution random variable with mean 23 and standard deviation 20. This value was set according to a number of studies. Those studies show that Pair Programming reduces the Project time by a factor starting from 0 (no change) in some cases to 45% in others. More data regarding those studies is available in the Background section.

- Test Driven development Impact Factor (*TDD_Impactfactor*): This factor represents the impact of adopting Test Driven Development practices in the team velocity. This value was set to a normal distribution random variable with mean -32 and standard deviation 42. This value was calculated according to number of studies [4]. Those studies show that Test Driven Development results in increasing the project time by a factor starting from 0 (no increase) in some studies to 80% in others. A review of those studies is available in the Background section.

- Team Velocity: This value represents the number of story points the team can develop per day taking into account all the previous factors. Equation 2 represents this value.

   *Team Velocity = Team_initial_velocity* (1+PP_usage*PP_ImpactFactor/100)*
   *(1+TDD_usage*TDD_impactfactor/100)*                          Equation (2)

## 3.3. Defected Story Points Model

This model calculates an estimate number for the defected story points to be re-developed in the next release. This number is affected by two XP practices: Test Driven development and Onsite Customer practices. Different components of the model are described as follows:

- Dev. Productivity: The developer productivity measured as the number Line Of Code (LOC) per day. According to the literature [4], a normal distribution with mean 40 and Standard Deviation of 20 represents this value.

- Estimated Release KLOC: represents the number of KLOC produced from this release. This value is calculated as the product of multiplying *Dev. Productivity* times *Team size* times *Estimated Release Days.*

- Defect Injection Ratio: represents the number of defects per KLOC. This value was set to a normal distribution with mean 20 and standard deviation 5 [4].

- Defect Rate: represents the number of defects in this release. It is calculated as the multiplication of the *Estimated Release KLOC* times *Defect Injection Ratio.*

- Defected Story Points: This value represents the number of defected story points to be re-developed in the next release taking into account the impact of two XP practices: Test Driven development and Onsite Customer practices (Equation 3). *OSC_Impact_Factor* and *TDD_Impact_Factor* represent the impact of the Onsite Customer and Test Driven development practices on reducing the defect rate. According to the literature, there values were set to 0.8 and 0.4 respectively [3],[4]. More details regarding the impact of these practices in the defect rate are available in the Background section.

*Defected_Story_Points = Defect_Rate*(1- OSC_Impact_Factor * onsitecustomer_usage )*(1 TDD_Impact_Factor *tddusage)*        Equation (3)

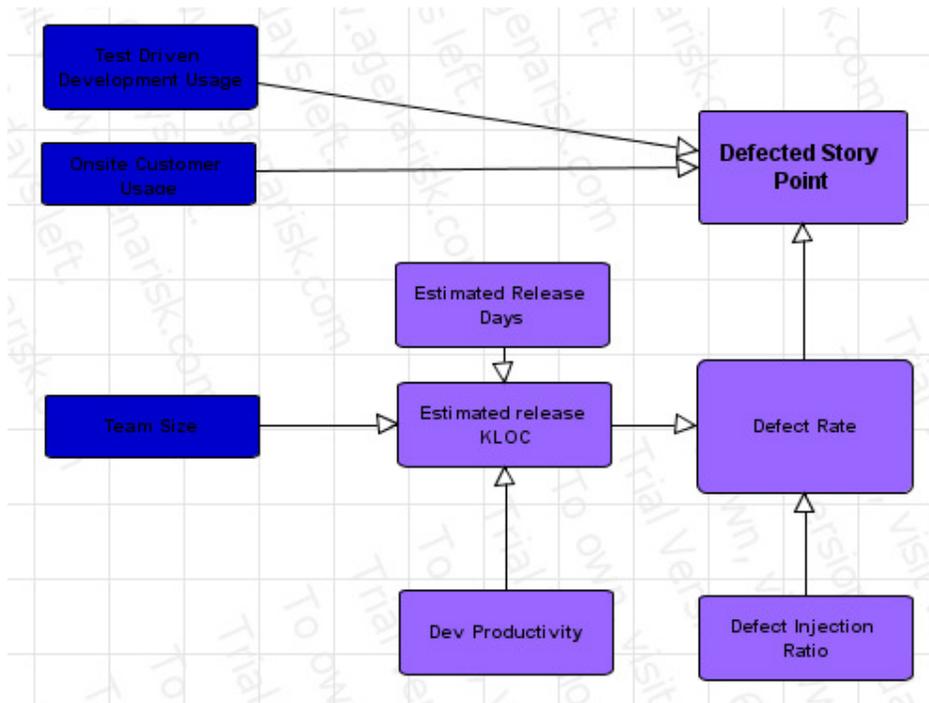

Figure 5 Defected Story Points Model

# 4. MODEL VALIDATION

The proposed model was implemented using AgenaRisk toolset [1]. AgenaRisk is a powerful tool for modelling risk and making predictions based on Bayesian Network. AgenaRisk has the following features:

- It integrates the advantages of Bayesian Networks, statistical simulation and spreadsheet analysis.

- A wide range of built-in conditional probability functions are available.

- It has the ability to build dynamic models.

- AgenaRisk is visual, simple and powerful tool.

A free licence for AgenaRisk toolset is available through the company website (http://www.agenarisk.com), but limited to 7 days.

In the next section, experiments setup will be illustrated, while the results will be provided and discussed in the following section

## 4.1.  Experiments Setup

Collecting data from real projects to validate our model was a difficult task due to several reasons. Due to XP *simplicity* value, it is difficult to find company collecting information regarding their activities and practices. Moreover, most real XP projects are developed by private companies having restrictions on publishing their internal development process. In addition, there is no guarantee that the available data is sufficient for model validation.

Two XP projects provided enough data to test our model. The first one is the *Repo Margining System* project [4]. The second one is a controlled case study reported by Pekka Abrahamsson [17]. We will refer to this case study in the rest of this paper by *Abrahamsson Case Study*. The model input data for the two projects are shown in tables 2 and 3. The model internal parameters are summarized in table 4.

Table 2 Repo Margining System input data

| *Repo Margining System* | Release 1 | Release 2 |
|---|---|---|
| Planned User Stories | 15 | 14 |
| Average Story Points per User Story | 15 | 15 |
| Team Size | 4 | 4 |
| Pair Programming usage | *About half* | *About half* |
| Test Driven Development usage | *About half* | *About half* |
| Onsite Customer usage | *Never* | *Never* |

Table 3 Abrahamsson Case Study input data

| Abrahamsson Case Study | Release 1 | Release 2 |
|---|---|---|
| Planned User Stories | 5 | 8 |
| Average Story Points per User Story | 10 | 8 |
| Team Size | 4 | 4 |
| Pair Programming usage | Almost Used | Frequently |
| Test Driven Development usage | About half | About half |
| Onsite Customer usage | Occasionally | Occasionally |

Table 4 Model internal parameters (U(a,b) refers to uniform distribution from a to b, while N(μ,σ) refers to normal distribution with mean μ and standard deviation σ)

| Random Variable | Value |
|---|---|
| Developer Initial Skills | U(1,10) |
| Developer Initial Velocity | N (4,1) |
| PP_Impact_Factor | N(23,20) |
| TDD_Impact_Factor | N(-32,42) |
| Dev. Productivity (LOC/Day) | N(40,20) |
| Defect Injection Ratio (defects/KLOC) | N(20,5) |

## 4.2.  Results and Discussion

Table 5 shows a comparison between the Estimated and the Real values for the two projects. Regarding the number of days, the estimated number is so close to the Real project. This indicates the acceptable accuracy of the proposed Team Velocity Model. On the other hand, the *Defected Story Points* model was not that accurate. For *Abrahamsson Case Study,* the estimated number of *defected story points* was close to the actual one, while for the *Repo Margining System,* the accuracy of the prediction system was not that good. The inaccuracy also appears in estimated the produced number of Line of codes for *Abrahamsson Case Study.*

The imprecision in some of the results, especially in the *Defected Story Points* model, is due to fixing some variables that should not be fixed. For example, the *Defect Injection Ratio* was fixed for the two projects to follow the normal distribution with mean 20 defects per KLOC. This value differs from project to another and should not be fixed. The same for *Developer Productivity* random variable that set to follow the normal distribution with mean 40 lines/day. This also depends on the nature of the project and should vary from project to another.

Table 5 Comparison between the Experiment Results and Real Project

|  | *Repo Margining System* | | *Abrahamsson Case Study* | |
|---|---|---|---|---|
|  | Experiment Results | Real Project [4] | Experiment Results | Real Project [17] |
| Number of Days | 65 | 60 | 11 | 12 |
| Defected Story Points | 200 | 319 | 11 | 9 |
| Lines of Code (KLOC) | 8.6 | 9.8 | 1.3 | 4.2 |

One solution of such imprecision is to adopt the model for self-learning, by which the model can learn from the first iterations and adjust different parameters and variables. This increases the confidence of the prediction and can correct the model's prior assumptions. This learning capability is a good extension for the proposed model.
Figures 6 and 7 show the estimated project status as time passes. Those curves gave accurate estimated values for the project finish time. Those curves can be obtained in the project planning phase before starting the actual development using very simple input data. Using such curves, the success or the failure of the project can be detected in early stage.

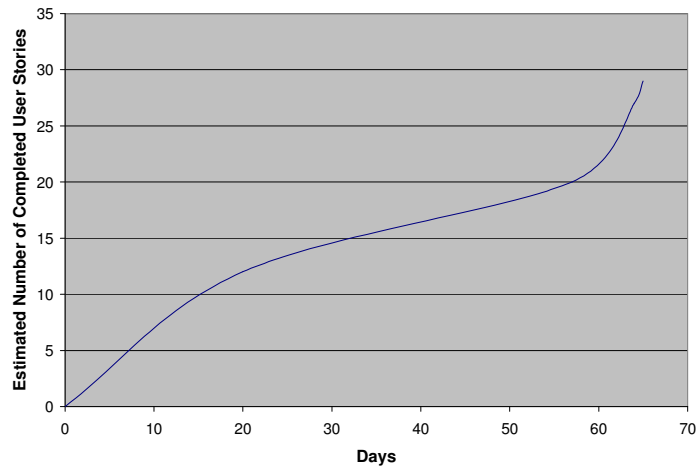

Figure 6 Repo Margining System project status curve

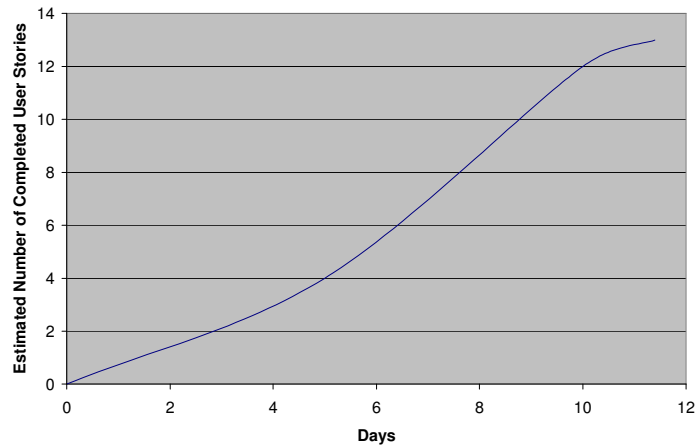

Figure 7 Abrahamsson Case Study project status curve

## 5. CONCLUSIONS

In this paper, a *Bayesian Network* based mathematical model for XP process is presented. The model can be used to predict the success/ failure of any XP project by estimating the expected finish time and the expected defect rate for each XP release. The proposed model comprises two internal models: *Team velocity* and *Defected Story Points* models. The model takes into account the impact of XP practices.

Two case studies were used for the validation of our model, namely: *Repo Margining System* and *Abrahamsson Case Study*. The results show that the model can be used successfully to predict the project finish time with a reasonable accuracy in the project planning phase using very simple input data. In addition, the results show that the accuracy of the *Team Velocity Model* is acceptable, while the *Defected Story Points* model was not that accurate.

Adopting the model to have a self-learning capability is a good extension of this work and can solve the imprecision in some of the results, especially in the *Defected Story Points* model, by which the model can learn from the first iterations and adjust different parameters and variables. This increases the confidence of the prediction and can correct the model's prior assumptions. This learning capability is a good extension for the proposed model.

**Authors**

**Mohamed Abouelela** received his M.Sc. degree in Engineering Mathematic from Cairo University, Cairo, Egypt in 2007. He is now a research assistant and Ph.D. candidate affiliated with the Software System Engineering Department at University of Regina, Regina, Canada. His main interests include Software Process modelling, XP assessment, Data management in grid systems, and optical grid networks.

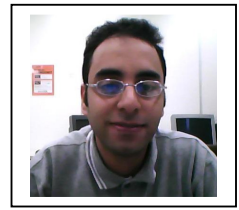

**Dr. Luigi Benedicenti** is a professor and Associate Dean in the Faculty of Engineering at the University of Regina. Benedicenti received his Laurea in Electrical Engineering and Ph.D. in Electrical and Computer Engineering from the University of Genoa, Italy. His collaborative network extends beyond Saskatchewan with TRLabs and IEEE, and Canada through collaborative work with colleagues in Europe, South East Asia, and North America.

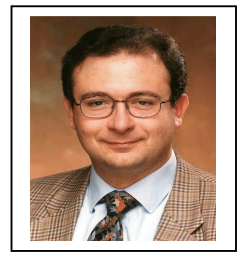